\documentstyle[12pt]{article}

\def\a{\alpha} \def\b{\beta} \def\e{\epsilon}
\def\d{\delta}
\begin{document}
\thispagestyle{empty}
\begin{flushright}
MPI-Ph/92-99\\ October 1992
\end{flushright}
\vskip 1cm
\bigskip\bigskip\begin{center} {\bf
\Huge{Jordanian Solutions of Simplex Equations}}
\end{center}  \vskip 2.0truecm
\centerline{\bf H. Ewen
and O. Ogievetsky${}^{\dag}$}
\vskip10mm
\centerline{Max-Planck-Institut f\"{u}r
 Physik, Werner-Heisenberg-Institut}
\centerline{F\"ohringer Ring 6, 8000 Munich 40, West Germany}

\vskip 6cm
\bigskip \nopagebreak \begin{abstract}
\vskip 0.4cm

\noindent
We construct for all $N$
a solution of the Frenkel--Moore $N$--simplex equation which
generalizes the $R$--matrix for the Jordanian quantum group.
\end{abstract}
\vskip 3.0cm
\bigskip \nopagebreak \begin{flushleft} \rule{2 in}{0.03cm}
\\ {\footnotesize \ ${}^{\dag}$ On leave from P.N. Lebedev Physical
 Institute, Theoretical Department, 117924 Moscow, Leninsky
 prospect 53, Russia}
 \end{flushleft}

\newpage\setcounter{page}1
\baselineskip 2em

\newcommand{\be}{\begin{equation}} \newcommand{\ee}{\end{equation}}

\newcommand{\beq}{\begin{equation}} \newcommand{\eeq}{\end{equation}}
\newcommand{\beqn}{\begin{displaymath}}
\newcommand{\eeqn}{\end{displaymath}}
\newcommand{\bea}{\begin{eqnarray*}} \newcommand{\eea}{\end{eqnarray*}}
\newcommand{\bean}{\begin{eqnarray}} \newcommand{\eean}{\end{eqnarray}}
\newcommand{\RR}{I\!\!R\,} \newcommand{\CC}{C\!\!\!\! I}
\newcommand{\skipline}{\vspace{\baselineskip}\newline}
\newcommand{\mand}{\mbox{ and }}
\newcommand{\ot}{\otimes}
\newcommand{\th}[1]{#1-th}

\section{Introduction}
The quantum Yang--Baxter, or
triangle equation can be included as a first member
in a sequence of simplex equations. There are several ways of defining
higher simplex equations. In this letter we investigate a version of
an $N$-simplex equation suggested in \cite{fm}. This is an equation for
an operator in the $N$-th power $V^{\otimes N}$ of a space $V$, that
is, for a tensor $R^{\a_1\ldots\a_N}_{\b_1\ldots\b_N}$.
Denote by $R_i$ the operator
\be (R_i)^{\a_1\ldots\a_{N+1}}_{\b_1\ldots\b_{N+1}}=
R^{\a_1\ldots\hat\a_i\ldots\a_{N+1}}_{\b_1\ldots\hat\b_i\ldots\b_{N+1}}
\d^{\a_i}_{\b_i}\ ,
\label{1}\ee
where the hat over an index means that this index is omitted.
The quantum $N$--simplex equation of \cite{fm} has the form
\be R_1 R_2 \ldots R_{N+1} = R_{N+1} R_N \ldots R_1 \ .
\label{2}\ee
For $N=2$ one recovers the standard Yang-Baxter equation.

The unit tensor satisfies (\ref{2}). For $R$ a  deformation of
the unit solution with parameter $\hbar$,
$R=1+\hbar r+{\cal O}(\hbar^2)$, the expansion of (\ref{2}) gives
in the second order in $\hbar$ the corresponding classical
$N$-simplex equation
\be \sum_{i<j}^{N+1} [r_i,r_j] = 0 \ ,\label{3}\ee
and $R$ is called a quantization of a classical solution $r$.

Solutions of the quantum Yang--Baxter equation give rise to matrix
quantum groups. For ${\rm dim} \ V=2$ there are
two types of SL quantum groups
\cite{ma,eow}. One is the standard SL$_q(2)$, another one is
the Jordanian quantum group SL$^J(2)$.

Frenkel and Moore \cite{fm} investigated the $N$-simplex generalization
of the $R$-matrix giving rise to SL$_q(2)$. They found that
the corresponding solutions
of the classical $N$-simplex equations exist for
all $N$. It turned out however that
the solutions do not admit a quantization for $N\geq 4$.

In this Letter we investigate an $N$-simplex
generalization of the Jordanian
$R$-matrix. The classical Jordanian $r$--matrix is a limiting case of the
nondegenerate SL$_q(2)$ $r$--matrix. The interesting property is that it
can be quantized for any $N$,
giving rise to nontrivial solutions of the quantum $N$-simplex equation.
The quantum Jordanian $R$--matrix turns out simply to be the
exponent of the classical one.

\section{Jordanian solution of the classical simplex equation}
The Jordanian deformation of SL(2) is induced by the
deformation of the Borel subalgebra \cite{o}.
In the matrix form, the classical
Jordanian $r$--matrix, $r=h\otimes e -e\otimes h$,
includes only two generators $h$, $e$ of
$sl(2)$, $h = \left( \begin{array}{cc} 1&0\\0&-1 \end{array}\right)$,
         $e = \left( \begin{array}{cc} 0&1\\0&0  \end{array}\right)$.
We will need only the algebraic properties of these matrices:
$he=-eh=e,\ e^2=0,\ h^2=1$.
Define operators acting in $V^{\otimes N}$:
\bean h_n^N &=& 1 \ot\ldots\ot h \ot\ldots\ot 1\ \ \
             \hbox{($h$ in the \th{n} position)} \nonumber\\
     e_n^N &=& (-1)^n 1 \ot\ldots\ot e \ot\ldots\ot 1\ \ \
     \hbox{($e$ in the \th{n} position)} \\
     H^N &=& h \ot h \ot\ldots\ot h\ . \nonumber \label{hnen}\eean
We will use these definitions for different values of $N$.
When there is no possibility of confusion we omit the superscript $N$.
We call an operator
\beq r=H \sum_{n=1}^N e_n\ \label{cr}\eeq
a Jordanian classical $r$--matrix.

\def\u{^}
Inserting the identity in the $i$-th place turns
an operator $e_n^N$ into
$e_n^{N+1}$ or $(-e_{n+1}^{N+1})$
depending on whether $n<i$ or $n\geq i$.
We have
\beq r_i = h_i\u{N+1} H\u{N+1}
\sum_{n=1}^{N+1} \sigma_{in} e_n\u{N+1}\ ,\ee
where $\sigma_{ij}=-1,0,+1$ for $i<j,i=j,i>j$ respectively.

The r-matrix (\ref{cr}) satisfies the classical $N$-simplex equation:
\beq \sum_{i<j}^{N+1} [r_i,r_j]
= -\sum_{i<j}^{N+1} \sum_{m,n}^{N+1} \sigma_{in}\sigma_{jm}[h_ie_n,h_je_m]
=- \sum_{i\neq j}^{N+1} \sum_{n}^{N+1}2\sigma_{in} h_je_ie_n = 0\ . \eeq

\section{Quantization}

For quantum groups, the explicit formulas for deformation of Lie
bialgebras \cite{d} to the third order in $\hbar$ implies that
$R=e^{\hbar r}=1+\hbar r+{1\over2} \hbar^2 r^2+...$ satisfies the quantum
Yang-Baxter equation up to order $\hbar^3$.
A direct calculation shows that this holds for simplex equations as well:
if $r$ satisfies
the classical simplex equation (\ref{3}),
then
$R=e^{\hbar r}$
satisfies the quantum simplex
equation (\ref{2}) up to order $\hbar^3$.

We are going to prove that for the Jordanian solution the exponent of
the classical $r$ solves the quantum Yang-Baxter
equation to all orders in $\hbar$.

The elements
$H e_n$ commute. Since they are nilpotent,
the $R$-matrix is actually a polynomial in $\hbar$,
\beq R=e^{\hbar r}=
e^{\hbar H\sum_{n=1}^N e_n}=\prod_{n=1}^N e^{\hbar H e_n}
= \prod_{n=1}^N (1+\hbar H e_n)\ .\label{rr}\eeq
Put
\be E_n=\hbar He_n\ .\ee
The elements $E_n$, $h_n$ satisfy the same algebra as
$e_n$, $h_n$: $h_nE_n=-E_nh_n=E_n$; $h_nE_m=E_mh_n$, $n\neq m$;
$E_nE_m=E_mE_n$, $n\neq m$; $E_n^2=0$.

In this  notation,
\beq R_i=e^{\hbar r_i}=
\prod_{n=1}^{N+1} (1+ h_i\u{N+1} \sigma_{in} E_n\u{N+1})\ .\eeq

We first prove a preparatory Lemma.

\vskip 0.3cm
\noindent{\bf Lemma.}
The operator $\prod_{n=1}^{N+1}(1+E_n)$ commutes with
$R_1 \ldots R_{N+1}$,
\beq [\prod_{n=1}^{N+1}(1+E_n)\ ,\ R_1 R_2 \ldots R_{N+1}] = 0
\ .\label{lem}\eeq

\noindent{\bf Proof.}
Denote the result of moving $E_k$ through
$ R_1\ldots R_{N+1}$ by $C_k$,
\beq E_k  R_1\ldots R_{N+1} =
           R_1\ldots R_{N+1} C_k\ .\eeq
We will derive a recursion formula for $C_k$ and
prove that $\prod_k(1+C_k) = \prod_k(1+E_k)$.

We have
\beq E_i (1+h_i\sigma_{in}E_n)
= (1+h_i\sigma_{in}E_n) E_i (1-2\sigma_{in}E_n)\ .\eeq
The element   $1-2\sigma_{in}E_n$ commutes
with each factor in $R_i$, therefore
\beq E_i  R_i
=  R_i E_i \prod_n(1-2\sigma_{in}E_n)\ .\label{er} \eeq
For $i\neq j$ we have
\be E_i  R_j= R_j E_i\ .\label{eirj}\ee
Moving $E_i$ through $R_1...R_{N+1}$ and using (\ref{er}), (\ref{eirj}),
we obtain immediately a recursion relation
\beq C_k=E_k \prod_{i<k}(1-2E_i)\prod_{j>k}(1+2C_j) \ .\label{ck}\eeq
Since $C_k$ is proportional to $E_k$, we have $C_k\u{2}=0$, and
$(1+2C_j)\u{-1}=1-2C_j$. Multiplying (\ref{ck}) by
$\prod_{j\leq k}(1+2C_j)$, we find
\beq C_k
\prod_{j<k}(1+2C_j)
=E_k\prod_{i<k}(1-2E_i)D\ ,
\label{ce}\eeq
where $D = \prod_{j=1}\u{N+1} (1+2C_j)$.
Summing over $k$, and using easily checked identities
\be 2\sum_kC_k\prod_{j<k}(1+2C_j)=\prod_k(1+2C_k)-1\ ,\
 2\sum_kE_k\prod_{j<k}(1-2E_j)=1-\prod_k(1-2E_k)\ ,\label{iden}\ee
we obtain $D\u{-1}=\prod (1-2E_j)$, or, taking inverses,
$D=\prod (1+2E_j)$. For nilpotent $\e$  the
expansion $(1+\e )\u{1/2}=1+1/2 \e -1/8 \e\u{2}+...$ terminates.
Using  this expression for the
square root of both sides of $D=\prod (1+2E_j)$,
we find
\be \prod_k (1+C_k)=\prod_k (1+E_k)\ .\ee
The proof is finished.

\vskip 0.5cm

\noindent{\bf Theorem.}
The $R$-matrix (\ref{rr}) solves
the quantum N-simplex equation (\ref{2}).
\vskip 0.2cm

\newcommand{\XN}{X^N}
\newcommand{\XNP}{X^{N+1}}
\noindent
{\bf Proof.} Using
$H (1+\hbar H e_n)=(1-\hbar H e_n) H$ and $H\u{2}=1$
we find $R^{-1}=HRH$ for the $R$-matrix (\ref{rr}). Therefore
we can rewrite the quantum $N$--simplex equation in the form
\beq (H R_1 R_2 \ldots R_{N+1})^2 = 1 \ .\label{hr}\eeq
Decomposing $H$ into the product of $h_i$,
$H=\prod_{i=1}^{N+1} h_i$ and moving every
$h_i$ to the corresponding $R_i$,
we rewrite the simplex equation once more:
\beq (\XNP_1 \XNP_2 \ldots \XNP_{N+1})^2 = 1 \ ,\label{xx}\eeq
where
\beq \XNP_i=h_i \prod_{n\neq i}^{N+1} (1+  h_iE_n)\eeq

\newcommand{\eN}{E_{N+1}}\newcommand{\hN}{h_{N+1}}
We will prove (\ref{xx}) by induction in $N$.
For $N=1$ the equation is easily verified.
Now we expand $\XNP_i=(\XN_i\ot 1)(1+h_iE_{N+1})$, $i<N+1$.
We will write $\XN_i$ instead of $\XN_i\ot 1$.
Then
\be (\XNP_1 \XNP_2 \ldots \XNP_{N+1})^2 = (\ \XN_1(1+h_1\eN) \ldots
\XN_N(1+h_N\eN) \hN \prod_{n=1}^N(1+\hN E_n)\ )^2\ ,\ee
which after moving $h_{N+1}$ to the right and using $h_{N+1}\u{2}=1$
gives
\bea
& & \XN_1(1+h_1\eN) \ldots \XN_N(1+h_N\eN) \prod_{n=1}^N(1+\hN E_n)\\
& & \times\ \XN_1(1-h_1\eN)
\ldots \XN_N(1-h_N\eN) \prod_{m=1}^N(1+\hN E_m)\ .
\eea
We have
\beqn \XN_i (1+h_j\eN) = (1-2h_iE_j\eN) (1+h_j\eN) \XN_i\ .\eeqn
Therefore
\bea \lefteqn{ \XN_1(1+h_1\eN) \ldots \XN_N(1+h_N\eN) }\\
=&&\!\!\prod_{i=1}^N (1+h_i\eN) \prod_{i_1<i_2}^N (1-2h_{i1}E_{i2}\eN)
\!  \prod_{i_1<i_2<i_3}^N (1+4h_{i1}E_{i2}E_{i3}\eN)\ldots
\XN_1\!\!\ldots\XN_N\\
=&&\!\!\prod_{k=1}^N [1+h_k\prod_{i>k}^N(1-2E_i)\eN]\
\XN_1 \ldots \XN_N\ .
\eea
Analogously,
\bea \lefteqn{ \XN_1(1-h_1\eN) \ldots \XN_N(1-h_N\eN) }\\
&=& \XN_1 \ldots \XN_N\
\prod_{k=1}^N [1-h_k\prod_{i<k}^N(1+2E_i)\eN]\ .\eea
Hence
\bean (\XNP_1\ldots \XNP_{N+1})^2 &=&
    \prod_{k=1}^N [1+h_k\prod_{i>k}^N(1-2E_i)\eN]\ \XN_1 \ldots \XN_N
\prod_{n=1}^N(1+E_n\hN) \nonumber \\
&\times&\!\!\!\XN_1 \ldots
\XN_N \prod_{l=1}^N [1-h_l\prod_{j<l}^N(1+2E_j)\eN]
\prod_{m=1}^N(1+E_m\hN).\label{zu}\eean
By the Lemma, $R_1...R_{N+1}$ commutes with $\prod (1+E_n)$, and
therefore with $(\prod (1+E_n))\u{-1}=\prod (1-E_n)$ as well.
For the operators $\XNP$ this amounts to
\beq \prod_{n=1}^{N+1}(1\pm E_n)\ \XNP_1 \ldots \XNP_{N+1} =
\XNP_1 \ldots \XNP_{N+1} \prod_{m=1}^{N+1}(1\mp E_m)\ .\label{xcom}\eeq
Since the operators $\XN_i$ do not touch the $(N+1)$-th space, and
$h_{N+1}$ takes values $\pm 1$ on the $(N+1)$-th space, we can rewrite
(\ref{xcom}) for $N$ in the form
\beq \prod_{n=1}^{N}(1+ E_nh_{N+1})\ \XN_1 \ldots \XN_N =
\XN_1 \ldots \XN_N\prod_{m=1}^{N}(1-E_mh_{N+1})\ .\label{xm}\eeq
Using (\ref{xm}) and the induction assumption for $N-1$ we find that
the right hand side of (\ref{zu}) equals
$$ \prod_{k=1}^N [1+
h_k\prod_{i>k}^N(1-2E_i)\eN]\ \prod_{n=1}^N(1-E_n\hN)
$$ \be \times
\prod_{l=1}^N [1-h_l\prod_{j<l}^N(1+2E_j)\eN]\
\prod_{m=1}^N(1+E_m\hN)\ .\label{wu}\ee
Finally,
\bean \lefteqn{ \prod_{n=1}^N(1-E_n\hN)
\ \prod_{l=1}^N [1-h_l\prod_{j<l}^N(1+2E_j)\eN]\
\prod_{m=1}^N(1+E_m\hN)}\nonumber\\
&=& \prod_{n=1}^N(1-E_n\hN)
\ [1-\sum_{l=1}^N h_l\prod_{j<l}^N(1+2E_j)\eN] \
\prod_{m=1}^N(1+E_m\hN)\nonumber\\
&=& 1- \prod_{n=1}^N(1-E_n)
 ( \sum_{l=1}^N h_l\prod_{j<l}^N(1+2E_j)\eN ) \prod_{m=1}^N(1-E_m)
\nonumber \\
&=& 1- \sum_{l=1}^N h_l\prod_{j\leq l}^N(1+2E_j)
\prod_{n=1}^N(1-E_n)  \eN
\prod_{m=1}^N(1-E_m)\nonumber\\
&=& 1-\sum_{l=1}^N h_l\prod_{j>l}^N(1-2E_j)\eN \nonumber\\
&=& \prod_{l=1}^N [1 - h_l\prod_{j>l}^N(1-2E_j)\eN]\ .\label{lst}\eean
In the second equality we used $\hN\eN=-\eN\hN=\eN$. The element
$h_l$ does not commute with $E_l$, which changes the summation
over $j<l$ to $j\leq l$ in the third equality.

Therefore, expression
(\ref{wu}) equals 1, which finishes the proof of the
Theorem.

\section*{Acknowledgements}
We are grateful to V.Jain, W.Schmidke, M.Schlieker and J. Wess
for various discussions.

\end{document}